\documentstyle[preprint,aps]{revtex}

\begin{document}
\author{{\it E. J. Ferrer and V. de la Incera}}
\address{Dept. of Physics, State University of New York, Fredonia, NY 14063, USA}
\title{BOUNDARY EFFECTS IN 2+1 DIMENSIONAL MAXWELL-CHERN-SIMONS THEORY}
\maketitle

\begin{abstract}
The effects of the sample's boundaries in the magnetic response of the
charged anyon fluid at finite temperature are investigated. For the case of
an infinite-strip sample it is shown that the Meissner effect takes place at
temperatures lower than the fermion energy gap $\omega _{c}$. The
temperature dependence of the corresponding effective penetration depth is
determined. At temperatures much larger than the scale $\omega _{c}$, a
different phase is found, in which the external magnetic field penetrates
the fluid.
\end{abstract}

\newpage

\section{Introduction}

Recently much attention has been given to lower dimensional gauge theories.
Such remarkable results as the chiral symmetry breaking \cite{1}, quantum
Hall effect \cite{2}, spontaneously broken Lorentz invariance by the
dynamical generation of a magnetic field \cite{3}, and the connection
between non-perturbative effects in low-energy strong interactions and QCD$%
_{2}$ \cite{4}, show the broad range of applicability of these theories.

In particular, 2+1 dimensional gauge theories with fractional statistics
-anyon systems \cite{4a}- have been extensively studied. One main reason for
such an interest has been the belief that a strongly correlated electron
system in two dimensions can be described by an effective field theory of
anyons \cite{5}, \cite{5a}. Besides, it has been claimed that anyons could
play a basic role in high-T$_{C}$ superconductivity \cite{5a}-\cite{6b}. It
is known \cite{a} that a charged anyon system in two spatial dimensions can
be modeled by means of a 2+1 dimensional Maxwell-Chern-Simons (MCS) theory.
An important feature of this theory is that it violates parity and
time-reversal invariance. However, at present no experimental evidences of P
and T violation in high-T$_{C}$ superconductivity have been found. It should
be pointed out, nevertheless, that it is possible to construct more
sophisticated P and T invariant anyonic models\cite{6a}. In any case,
whether linked to high-T$_{C}$ superconductivity or not, the anyon system is
an interesting theoretical model in its own right.

The superconducting behavior of anyon systems at $T=0$ has been investigated
by many authors \cite{6}-\cite{15a}. Crucial to the existence of anyon
superconductivity at $T=0$ is the exact cancellation between the bare and
induced Chern-Simons terms in the effective action of the theory.

Although a general consensus exists regarding the superconductivity of anyon
systems at zero temperature, a similar consensus at finite temperature is
yet to be achieved \cite{8}-\cite{16}. Some authors (see ref. \cite{9}) have
concluded that the superconductivity is lost at $T\neq 0$, based upon the
appearance of a temperature-dependent correction to the induced Chern-Simons
coefficient that is not cancelled out by the bare term. In ref. \cite{11} it
is argued, however, that this finite temperature correction is numerically
negligible at $T<200$ $K$, and that the main reason for the lack of a
Meissner effect is the development of a pole $\sim \left( \frac{1}{{\bf k}%
^{2}}\right) $ in the polarization operator component $\Pi _{00}$ at $T\neq
0 $. There, it is discussed how the existence of this pole leads to a so
called partial Meissner effect with a constant magnetic field penetration
throughout the sample that appreciably increases with temperature. On the
other hand, in ref. \cite{8}, it has been independently claimed that the
anyon model cannot superconduct at finite temperature due to the existence
of a long-range mode, found inside the infinite bulk at $T\neq 0$. The long
range mode found in ref. \cite{8} is also a consequence of the existence of
a pole $\sim \left( \frac{1}{{\bf k}^{2}}\right) $ in the polarization
operator component $\Pi _{00}$ at $T\neq 0$.

The apparent lack of superconductivity at temperatures greater than zero has
been considered as a discouraging property of anyon models. Nevertheless, it
may be still premature to disregard the anyons as a feasible solution for
explaining high -T$_{c}$ superconductivity, at least if the reason
sustaining such a belief is the absence of the Meissner effect at finite
temperature. As it was shown in a previous paper \cite{16}, the lack of a
Meissner effect, reported in ref. \cite{11} for the case of a half-plane
sample as a partial Meissner effect, is a direct consequence of the omission
of the sample boundary effects in the calculations of the minimal solution
for the magnetic field within the sample. To understand this remark we must
take into account that the results of ref. \cite{11} were obtained by
finding the magnetization in the bulk due to an externally applied magnetic
field at the boundary of a half-plane sample. However, in doing so, a
uniform magnetization was assumed and therefore the boundary effects were
indeed neglected. Besides, in ref. \cite{11} the field equations were solved
considering only one short-range mode of propagation for the magnetic field,
while as has been emphasized in our previous letter \cite{16}, there is a
second short-range mode whose qualitative contribution to the solutions of
the field equations cannot be ignored.

In the present paper we study the effects of the sample's boundaries in the
magnetic response of the anyon fluid at finite temperature. This is done by
considering a sample shaped as an infinite strip. When a constant and
homogeneous external magnetic field, which is perpendicular to the sample
plane, is applied at the boundaries of the strip, two different magnetic
responses, depending on the temperature values, can be identified. At
temperatures smaller than the fermion energy gap inherent to the
many-particle MCS model ($T\ll \omega _{c}$), the system exhibits a Meissner
effect. In this case the magnetic field cannot penetrate the bulk farther
than a very short distance ($\overline{\lambda }\sim 10^{-5}cm$ for electron
densities characteristic of the high -T$_{c}$ superconductors and $T\sim 200$
$K$). On the other hand, as it is natural to expect from a physical point of
view, when the temperatures are larger than the energy gap ($T\gg \omega
_{c} $) the Meissner effect is lost. In this temperature region a periodic
inhomogeneous magnetic field is present within the bulk.

These results, together with those previously reported in ref. \cite{16},
indicate that, contrary to some authors' belief, the superconducting
behavior (more precisely, the Meissner effect), found in the charged anyon
fluid at $T=0$, does not disappear as soon as the system is heated.

As it is shown below, the presence of boundaries can affect the dynamics of
the system in such a way that the mode that accounts for a homogeneous field
penetration \cite{8} cannot propagate in the bulk. Although these results
have been proved for two types of samples, the half-plane \cite{16} and the
infinite strip reported in this paper, we conjecture that similar effects
should also exist in other geometries.

Our main conclusion is that the magnetic behavior of the anyon fluid is not
just determined by its bulk properties, but it is essentially affected by
the sample boundary conditions. The importance of the boundary conditions in
2+1 dimensional models has been previously stressed in ref.\cite{b}.

The plan for the paper is as follows. In Sec. 2, for completeness as well as
for the convenience of the reader, we define the many-particle 2+1
dimensional MCS model used to describe the charged anyon fluid, and briefly
review its main characteristics. In Sec. 3 we study the magnetic response in
the self-consistent field approximation of a charged anyon fluid confined to
an infinite-strip, finding the analytical solution of the MCS field
equations that satisfies the boundary conditions. The fermion contribution
in this approximation is given by the corresponding polarization operators
at $T\neq 0$ in the background of a many-particle induced Chern-Simons
magnetic field. Using these polarization operators in the low temperature
approximation ($T\ll \omega _{c}$), we determine the system's two London
penetration depths. Taking into account that the boundary conditions are not
enough to completely determine the magnetic field solution within the
sample, an extra physical condition, the minimization of the system
free-energy density, is imposed. This is done in Sec. 4. In this section we
prove that even though the electromagnetic field has a long-range mode of
propagation in the charged anyon fluid at $T\neq 0$ \cite{8}, a constant and
uniform magnetic field applied at the sample's boundaries cannot propagate
through this mode. The explicit temperature dependence at $T\ll \omega _{c}$
of all the coefficients appearing in the magnetic field solution, and of the
effective London penetration depth are also found. In Sec. 5, we discuss how
the superconducting behavior of the charged anyon fluid disappears at
temperatures larger than the energy gap ($T\gg \omega _{c}$). Sec. 6
contains the summary and discussion.

\section{MCS Many-Particle Model}

The Lagrangian density of the 2+1 dimensional non-relativistic charged MCS
system is

\begin{equation}
{\cal L}=-\frac{1}{4}F_{\mu \nu }^{2}-\frac{N}{4\pi }\varepsilon ^{\mu \nu
\rho }a_{\mu }\partial _{\nu }a_{\rho }+en_{e}A_{0}+i\psi ^{\dagger
}D_{0}\psi -\frac{1}{2m}\left| D_{k}\psi \right| ^{2}+\psi ^{\dagger }\mu
\psi  \eqnum{2.1}
\end{equation}
where $A_{\mu }$ and $a_{\mu }$ represent the electromagnetic and the
Chern-Simons fields respectively. The role of the Chern-Simons fields is
simply to change the quantum statistics of the matter field, thus, they do
not have an independent dynamics. $\psi $ represents non-relativistic
spinless fermions. $N\ $ is a positive integer that determines the magnitude
of the Chern-$%
\mathop{\rm Si}%
$mons coupling constant. The charged character of the system is implemented
by introducing a chemical potential $\mu $; $n_{e}$ is a background
neutralizing `classical' charge density, and $m$ is the fermion mass. We
will consider throughout the paper the metric $g_{\mu \nu }$=$(1,-%
\overrightarrow{1})$. The covariant derivative $D_{\nu }$ is given by

\begin{equation}
D_{\nu }=\partial _{\nu }+i\left( a_{\nu }+eA_{\nu }\right) ,\qquad \nu
=0,1,2  \eqnum{2.2}
\end{equation}

It is known that to guarantee the system neutrality in the presence of a
different from zero fermion density $\left( n_{e}\neq 0\right) $,$\ $a
nontrivial background of Chern-Simons magnetic field $\left( \overline{b}=%
\overline{f}_{21}\right) $ is generated. The Chern-Simons background field
is obtained as the solution of the mean field Euler-Lagrange equations
derived from (2.1)

\begin{mathletters}
\begin{equation}
-\frac{N}{4\pi }\varepsilon ^{\mu \nu \rho }f_{\nu \rho }=\left\langle
j^{\mu }\right\rangle  \eqnum{2.3}
\end{equation}

\begin{equation}
\partial _{\nu }F^{\mu \nu }=e\left\langle j^{\mu }\right\rangle
-en_{e}\delta ^{\mu 0}  \eqnum{2.4}
\end{equation}
considering that the system formed by the electron fluid and the background
charge $n_{e}$ is neutral

\end{mathletters}
\begin{equation}
\left\langle j^{0}\right\rangle -n_{e}\delta ^{\mu 0}=0  \eqnum{2.5}
\end{equation}
In eq. (2.5) $\left\langle j^{0}\right\rangle $ is the fermion density of
the many-particle fermion system

\begin{equation}
\left\langle j^{0}\right\rangle =\frac{\partial \Omega }{\partial \mu }, 
\eqnum{2.6}
\end{equation}
$\Omega $ is the fermion thermodynamic potential.

In this approximation it is found from (2.3)-(2.5) that the Chern-Simons
magnetic background is given by

\begin{equation}
\overline{b}=\frac{2\pi n_{e}}{N}  \eqnum{2.7}
\end{equation}

Then, the unperturbed one-particle Hamiltonian of the matter field
represents a particle in the background of the Chern-Simons magnetic field $%
\overline{b\text{,}}$

\begin{equation}
H_{0}=-\frac{1}{2m}\left[ \left( \partial _{1}+i\overline{b}x_{2}\right)
^{2}+\partial _{2}^{2}\right]  \eqnum{2.8}
\end{equation}
In (2.8) we considered the background Chern-Simons potential, $\overline{a}%
_{k}$, $(k=1,2)$, in the Landau gauge

\begin{equation}
\overline{a}_{k}=\overline{b}x_{2}\delta _{k1}  \eqnum{2.9}
\end{equation}

The eigenvalue problem defined by the Hamiltonian (2.8) with periodic
boundary conditions in the $x_{1}$-direction: $\Psi \left( x_{1}+L,\text{ }%
x_{2}\right) =$ $\Psi \left( x_{1},\text{ }x_{2}\right) $,

\begin{equation}
H_{0}\Psi _{nk}=\epsilon _{n}\Psi _{nk},\qquad n=0,1,2,...\text{ }and\text{ }%
k\in {\cal Z}  \eqnum{2.10}
\end{equation}
has eigenvalues and eigenfunctions given respectively by

\begin{equation}
\epsilon _{n}=\left( n+\frac{1}{2}\right) \omega _{c}\qquad  \eqnum{2.11}
\end{equation}

\begin{equation}
\Psi _{nk}=\frac{\overline{b}^{1/4}}{\sqrt{L}}\exp \left( -2\pi
ikx_{1}/L\right) \varphi _{n}\left( x_{2}\sqrt{\overline{b}}-\frac{2\pi k}{L%
\sqrt{\overline{b}}}\right)  \eqnum{2.12}
\end{equation}
where $\omega _{c}=\overline{b}/m$ is the cyclotron frequency and $\varphi
_{n}\left( \xi \right) $ are the orthonormalized harmonic oscillator wave
functions.

Note that the energy levels $\epsilon _{n}$ are degenerates (they do not
depend on $k$). Then, for each Landau level $n$ there exists a band of
degenerate states. The cyclotron frequency $\omega _{c}$ plays here the role
of the energy gap between occupied Landau levels. It is easy to prove that
the filling factor, defined as the ratio between the density of particles $%
n_{e}$ and the number of states per unit area of a full Landau level, is
equal to the Chern-$%
\mathop{\rm Si}%
$mons coupling constant $N$. Thus, because we are considering that $N$ is a
positive integer, we have in this MCS theory $N$ completely filled Landau
levels. Once this ground state is established, it can be argued immediately 
\cite{6}, \cite{6b}, \cite{10a}, \cite{15}, that at $T=0$ the system will be
confined to a filled band, which is separated by an energy gap from the free
states; therefore, it is natural to expect that at $T=0$ the system should
superconduct. This result is already a well established fact on the basis of
Hartree-Fock analysis\cite{6} and Random Phase Approximation \cite{6b},\cite
{10a}. The case at $T\neq 0$ is more controversial since thermal
fluctuations, occurring in the many-particle system, can produce significant
changes. As we will show in this paper, the presence in this theory of a
natural scale, the cyclotron frequency $\omega _{c}$, is crucial for the
existence of a phase at $T\ll \omega _{c}$, on which the system, when
confined to a bounded region, still behaves as a superconductor.

The fermion thermal Green's function in the presence of the background
Chern-Simons field $\overline{b}$

\begin{equation}
G\left( x,x^{\prime }\right) =-\left\langle T_{\tau }\psi \left( x\right) 
\overline{\psi }\left( x^{\prime }\right) \right\rangle  \eqnum{2.13}
\end{equation}
is obtained by solving the equation

\begin{equation}
\left( \partial _{\tau }-\frac{1}{2m}\overline{D}_{k}^{2}-\mu \right)
G\left( x,x^{\prime }\right) =-\delta _{3}\left( x-x^{\prime }\right) 
\eqnum{2.14}
\end{equation}
subject to the requirement of antiperiodicity under the imaginary time
translation $\tau \rightarrow \tau +\beta $ ($\beta $ is the inverse
absolute temperature). In (2.14) we have introduced the notation

\begin{equation}
\overline{D}_{k}=\partial _{k}+i\overline{a}_{k}  \eqnum{2.15}
\end{equation}

The Fourier transform of the fermion thermal Green's function (2.13)

\begin{equation}
G\left( p_{4},{\bf p}\right) =\int\limits_{0}^{\beta }d\tau \int d{\bf x}%
G\left( \tau ,{\bf x}\right) e^{i\left( p_{4}\tau -{\bf px}\right) } 
\eqnum{2.16}
\end{equation}
can be expressed in terms of the orthonormalized harmonic oscillator wave
functions $\varphi _{n}\left( \xi \right) $ as \cite{Efrain}

\begin{eqnarray}
G\left( p_{4},{\bf p}\right) &=&\int\limits_{0}^{\infty }d\rho
\int\limits_{-\infty }^{\infty }dx_{2}\sqrt{\overline{b}}\exp -\left(
ip_{2}x_{2}\right) \exp -\left( ip_{4}+\mu -\frac{\overline{b}}{2m}\right)
\rho  \nonumber \\
&&\sum\limits_{n=0}^{\infty }\varphi _{n}\left( \xi \right) \varphi
_{n}\left( \xi ^{\prime }\right) t^{n}  \eqnum{2.17}
\end{eqnarray}
where $t=\exp \frac{\overline{b}}{m}\rho $, $\xi =\frac{p_{1}}{\sqrt{%
\overline{b}}}+\frac{x_{2}\sqrt{\overline{b}}}{2}$, $\xi ^{\prime }=\frac{%
p_{1}}{\sqrt{\overline{b}}}-\frac{x_{2}\sqrt{\overline{b}}}{2}$ and $%
p_{4}=(2n+1)\pi /\beta $ are the discrete frequencies $(n=0,1,2,...)$
corresponding to fermion fields.

\section{Linear Response in the Infinite Strip}

\subsection{Effective Theory at $\mu \neq 0$ and $T\neq 0$}

In ref.\cite{8} the effective current-current interaction of the MCS model
was calculated to determine the independent components of the magnetic
interaction at finite temperature in a sample without boundaries, i.e., in
the free space. These authors concluded that the pure Meissner effect
observed at zero temperature is certainly compromised by the appearance of a
long-range mode at $T\neq 0$. Our main goal in the present paper is to
investigate the magnetic response of the charged anyon fluid at finite
temperature for a sample that confines the fluid within some specific
boundaries. As we prove henceforth, the confinement of the system to a
bounded region (a condition which is closer to the experimental situation
than the free-space case) is crucial for the realization of the Meissner
effect inside the charged anyon fluid at finite temperature.

Let us investigate the linear response of a charged anyon fluid at finite
temperature and density to an externally applied magnetic field in the
specific case of an infinite-strip sample. The linear response of the medium
can be found under the assumption that the quantum fluctuations of the gauge
fields about the ground-state are small. In this case the one-loop fermion
contribution to the effective action, obtained after integrating out the
fermion fields, can be evaluated up to second order in the gauge fields. The
effective action of the theory within this linear approximation \cite{8},%
\cite{11} takes the form

\begin{equation}
\Gamma _{eff}\,\left( A_{\nu },a_{\nu }\right) =\int dx\left( -\frac{1}{4}%
F_{\mu \nu }^{2}-\frac{N}{4\pi }\varepsilon ^{\mu \nu \rho }a_{\mu }\partial
_{\nu }a_{\rho }+en_{e}A_{0}\right) +\Gamma ^{\left( 2\right) }  \eqnum{3.1}
\end{equation}

\[
\Gamma ^{\left( 2\right) }=\int dx\Pi ^{\nu }\left( x\right) \left[ a_{\nu
}\left( x\right) +eA_{\nu }\left( x\right) \right] +\int dxdy\left[ a_{\nu
}\left( x\right) +eA_{\nu }\left( x\right) \right] \Pi ^{\mu \nu }\left(
x,y\right) \left[ a_{\nu }\left( y\right) +eA_{\nu }\left( y\right) \right] 
\]
Here $\Gamma ^{\left( 2\right) }$ is the one-loop fermion contribution to
the effective action in the linear approximation. The operators $\Pi _{\nu }$
and $\Pi _{\mu \nu }$ are calculated using the fermion thermal Green's
function in the presence of the background field $\overline{b}$ (2.17). They
represent the fermion tadpole and one-loop polarization operators
respectively. Their leading behaviors for static $\left( k_{0}=0\right) $
and slowly $\left( {\bf k}\sim 0\right) $ varying configurations in the
frame ${\bf k}=(k,0)$ take the form

\begin{equation}
\Pi _{k}\left( x\right) =0,\;\;\;\Pi _{0}\left( x\right) =-n_{e},\;\;\;\Pi
_{\mu \nu }=\left( 
\begin{array}{ccc}
{\it \Pi }_{{\it 0}}+{\it \Pi }_{{\it 0}}\,^{\prime }\,k^{2} & 0 & {\it \Pi }%
_{{\it 1}}k \\ 
0 & 0 & 0 \\ 
-{\it \Pi }_{{\it 1}}k & 0 & {\it \Pi }_{\,{\it 2}}k^{2}
\end{array}
\right) ,  \eqnum{3.2}
\end{equation}

The independent coefficients: ${\it \Pi }_{{\it 0}}$, ${\it \Pi }_{{\it 0}%
}\,^{\prime }$, ${\it \Pi }_{{\it 1}}$ and ${\it \Pi }_{\,{\it 2}}$ are
functions of $k^{2}$, $\mu $ and $\overline{b}$. In order to find them we
just need to calculate the $\Pi _{\mu \nu }$ Euclidean components: $\Pi
_{44} $, $\Pi _{42}$ and $\Pi _{22}$. In the Landau gauge these Euclidean
components are given by\cite{11},

\begin{mathletters}
\begin{equation}
\Pi _{44}\left( k,\mu ,\overline{b}\right) =-\frac{1}{\beta }%
\sum\limits_{p_{4}}\frac{d{\bf p}}{\left( 2\pi \right) ^{2}}G\left( p\right)
G\left( p-k\right) ,  \eqnum{3.3}
\end{equation}

\begin{equation}
\Pi _{4j}\left( k,\mu ,\overline{b}\right) =\frac{i}{2m\beta }%
\sum\limits_{p_{4}}\frac{d{\bf p}}{\left( 2\pi \right) ^{2}}\left\{ G\left(
p\right) \cdot D_{j}^{-}G\left( p-k\right) +D_{j}^{+}G\left( p\right) \cdot
G\left( p-k\right) \right\} ,  \eqnum{3.4}
\end{equation}

\end{mathletters}
\begin{eqnarray}
\Pi _{jk}\left( k,\mu ,\overline{b}\right) &=&\frac{1}{4m^{2}\beta }%
\sum\limits_{p_{4}}\frac{d{\bf p}}{\left( 2\pi \right) ^{2}}\left\{
D_{k}^{-}G\left( p\right) \cdot D_{j}^{-}G\left( p-k\right)
+D_{j}^{+}G\left( p\right) \cdot D_{k}^{+}G\left( p-k\right) \right. 
\nonumber \\
&&\left. +D_{j}^{+}D_{k}^{-}G\left( p\right) \cdot G\left( p-k\right)
+G\left( p\right) \cdot D_{j}^{-}D_{k}^{+}G\left( p-k\right) \right\} 
\nonumber \\
&&-\frac{1}{2m}\Pi _{4},  \eqnum{3.5}
\end{eqnarray}
where the notation

\begin{eqnarray}
D_{j}^{\pm }G\left( p\right) &=&\left[ ip_{j}\mp \frac{\overline{b}}{2}%
\varepsilon ^{jk}\partial _{p_{k}}\right] G\left( p\right) ,  \nonumber \\
D_{j}^{\pm }G\left( p-k\right) &=&\left[ i\left( p_{j}-k_{j}\right) \mp 
\frac{\overline{b}}{2}\varepsilon ^{jk}\partial _{p_{k}}\right] G\left(
p-k\right) ,  \eqnum{3.6}
\end{eqnarray}
was used.

Using (3.3)-(3.5) after summing in $p_{4}$, we found that, in the $k/\sqrt{%
\overline{b}}\ll 1$ limit, the polarization operator coefficients ${\it \Pi }%
_{{\it 0}}$, ${\it \Pi }_{{\it 0}}\,^{\prime }$, ${\it \Pi }_{{\it 1}}$ and $%
{\it \Pi }_{\,{\it 2}}$ are

\[
{\it \Pi }_{{\it 0}}=\frac{\beta \overline{b}}{8\pi {\bf k}^{2}}%
\sum_{n}\Theta _{n},\;\qquad {\it \Pi }_{{\it 0}}\,^{\prime }=\frac{2m}{\pi 
\overline{b}}\sum_{n}\Delta _{n}-\frac{\beta }{8\pi }\sum_{n}(2n+1)\Theta
_{n}, 
\]

\[
{\it \Pi }_{{\it 1}}=\frac{1}{\pi }\sum_{n}\Delta _{n}-\frac{\beta \overline{%
b}}{16\pi m}\sum_{n}(2n+1)\Theta _{n},\qquad {\it \Pi }_{\,{\it 2}}=\frac{1}{%
\pi m}\sum_{n}(2n+1)\Delta _{n}-\frac{\beta \overline{b}}{32\pi m^{2}}%
\sum_{n}(2n+1)^{2}\Theta _{n}, 
\]

\begin{equation}
\Theta _{n}=%
\mathop{\rm sech}%
\,^{2}\frac{\beta (\epsilon _{n}/2-\mu )}{2},\qquad \Delta _{n}=(e^{\beta
(\epsilon _{n}/2-\mu )}+1)^{-1}  \eqnum{3.7}
\end{equation}

The leading contributions of the one-loop polarization operator coefficients
(3.7) at low temperatures $\left( T\ll \omega _{c}\right) $ are

\begin{equation}
{\it \Pi }_{{\it 0}}=\frac{2\beta \overline{b}}{\pi }e^{-\beta \overline{b}%
/2m},\qquad {\it \Pi }_{{\it 0}}\,^{\prime }=\frac{mN}{2\pi \overline{b}}%
{\it \Lambda },\qquad {\it \Pi }_{{\it 1}}=\frac{N}{2\pi }{\it \Lambda }%
,\quad {\it \Pi }_{\,{\it 2}}=\frac{N^{2}}{4\pi m}{\it \Lambda },\qquad {\it %
\Lambda }=\left[ 1-\frac{2\beta \overline{b}}{m}e^{-\beta \overline{b}%
/2m}\right]  \eqnum{3.8}
\end{equation}
and at high temperatures $\left( T\gg \omega _{c}\right) $ are

\begin{equation}
{\it \Pi }_{{\it 0}}=\frac{m}{2\pi }\left[ \tanh \frac{\beta \mu }{2}%
+1\right] ,\qquad {\it \Pi }_{{\it 0}}\,^{\prime }=-\frac{\beta }{48\pi }%
\mathop{\rm sech}%
\!^{2}\!\,\left( \frac{\beta \mu }{2}\right) ,\qquad {\it \Pi }_{{\it 1}}=%
\frac{\overline{b}}{m}{\it \Pi }_{{\it 0}}\,^{\prime },\qquad {\it \Pi }_{\,%
{\it 2}}=\frac{1}{12m^{2}}{\it \Pi }_{{\it 0}}  \eqnum{3.9}
\end{equation}
In these expressions $\mu $ is the chemical potential and $m=2m_{e}$ ($m_{e}$
is the electron mass). These results are in agreement with those found in
refs.\cite{8},\cite{14}.

\subsection{MCS Linear Equations}

To find the response of the anyon fluid to an externally applied magnetic
field, one needs to use the extremum equations derived from the effective
action (3.1). This formulation is known in the literature as the
self-consistent field approximation\cite{11}. In solving these equations we
confine our analysis to gauge field configurations which are static and
uniform in the y-direction. Within this restriction we take a gauge in which 
$A_{1}=a_{1}=0$.

The Maxwell and Chern-Simons extremum equations are respectively,

\begin{equation}
\partial _{\nu }F^{\nu \mu }=eJ_{ind}^{\mu }  \eqnum{3.10a}
\end{equation}

\begin{equation}
-\frac{N}{4\pi }\varepsilon ^{\mu \nu \rho }f_{\nu \rho }=J_{ind}^{\mu } 
\eqnum{3.10b}
\end{equation}
Here, $f_{\mu \nu }$ is the Chern-Simons gauge field strength tensor,
defined as $f_{\mu \nu }=\partial _{\mu }a_{\nu }-\partial _{\nu }a_{\mu }$,
and $J_{ind}^{\mu }$ is the current density induced by the anyon system at
finite temperature and density. Their different components are given by

\begin{equation}
J_{ind}^{0}\left( x\right) ={\it \Pi }_{{\it 0}}\left[ a_{0}\left( x\right)
+eA_{0}\left( x\right) \right] +{\it \Pi }_{{\it 0}}\,^{\prime }\partial
_{x}\left( {\cal E}+eE\right) +{\it \Pi }_{{\it 1}}\left( b+eB\right) 
\eqnum{3.11a}
\end{equation}

\begin{equation}
J_{ind}^{1}\left( x\right) =0,\qquad J_{ind}^{2}\left( x\right) ={\it \Pi }_{%
{\it 1}}\left( {\cal E}+eE\right) +{\it \Pi }_{\,{\it 2}}\partial _{x}\left(
b+eB\right)  \eqnum{3.11b}
\end{equation}
in the above expressions we used the following notation: ${\cal E}=f_{01}$, $%
E=F_{01}$, $b=f_{12}$ and $B=F_{12}$. Eqs. (3.11) play the role in the anyon
fluid of the London equations in BCS superconductivity. When the induced
currents (3.11) are substituted in eqs. (3.10) we find, after some
manipulation, the set of independent differential equations,

\begin{equation}
\omega \partial _{x}^{2}B+\alpha B=\gamma \left[ \partial _{x}E-\sigma
A_{0}\right] +\tau \,a_{0},  \eqnum{3.12}
\end{equation}

\begin{equation}
\partial _{x}B=\kappa \partial _{x}^{2}E+\eta E,  \eqnum{3.13}
\end{equation}

\begin{equation}
\partial _{x}a_{0}=\chi \partial _{x}B  \eqnum{3.14}
\end{equation}
The coefficients appearing in these differential equations depend on the
components of the polarization operators through the relations

\[
\omega =\frac{2\pi }{N}{\it \Pi }_{{\it 0}}\,^{\prime },\quad \alpha =-e^{2}%
{\it \Pi }_{{\it 1}},\quad \tau =e{\it \Pi }_{{\it 0}},\quad \chi =\frac{%
2\pi }{eN},\quad \sigma =-\frac{e^{2}}{\gamma }{\it \Pi }_{{\it 0}},\quad
\eta =-\frac{e^{2}}{\delta }{\it \Pi }_{{\it 1}}, 
\]

\begin{equation}
\gamma =1+e^{2}{\it \Pi }_{{\it 0}}\,^{\prime }-\frac{2\pi }{N}{\it \Pi }_{%
{\it 1}},\quad \delta =1+e^{2}{\it \Pi }_{\,{\it 2}}-\frac{2\pi }{N}{\it \Pi 
}_{{\it 1}},\quad \kappa =\frac{2\pi }{N\delta }{\it \Pi }_{\,{\it 2}}. 
\eqnum{3.15}
\end{equation}

Distinctive of eq. (3.12) is the presence of the nonzero coefficients $%
\sigma $ and $\tau $. They are related to the Debye screening in the two
dimensional anyon thermal ensemble. A characteristic of this 2+1 dimensional
model is that the Debye screening disappears at $T=0$, even if the chemical
potential is different from zero. Note that $\sigma $ and $\tau $ link the
magnetic field to the zero components of the gauge potentials, $A_{0}$ and $%
a_{0}$. As a consequence, these gauge potentials will play a nontrivial role
in finding the magnetic field solution of the system.

\subsection{Field Solutions and Boundary Conditions}

Using eqs.(3.12)-(3.14) one can obtain a higher order differential equation
that involves only the electric field,

\begin{equation}
a\partial _{x}^{4}E+d\partial _{x}^{2}E+cE=0,  \eqnum{3.16}
\end{equation}
Here, $a=\omega \kappa $, $d=\omega \eta +\alpha \kappa -\gamma -\tau \kappa
\chi $, and $c=\alpha \eta -\sigma \gamma -\tau \eta \chi $.

Solving (3.16) we find

\begin{equation}
E\left( x\right) =C_{1}e^{-x\xi _{1}}+C_{2}e^{x\xi _{1}}+C_{3}e^{-x\xi
_{2}}+C_{4}e^{x\xi _{2}},  \eqnum{3.17}
\end{equation}
where

\begin{equation}
\xi _{1,2}=\left[ -d\pm \sqrt{d^{2}-4ac}\right] ^{\frac{1}{2}}/\sqrt{2a} 
\eqnum{3.18}
\end{equation}
When the low density approximation $n_{e}\ll m^{2}$ is considered (this
approximation is in agreement with the typical values $n_{e}=2\times
10^{14}cm^{-2}$, $m_{e}=2.6\times 10^{10}cm^{-1}$ found in high-T$_{C}$
superconductivity), we find that, for $N=2$ and at temperatures lower than
the energy gap $\left( T\ll \omega _{c}\right) $, the inverse length scales
(3.18) are given by the following real functions 
\begin{equation}
\xi _{1}\simeq e\sqrt{\frac{m}{\pi }}\left[ 1+\left( \frac{\pi ^{2}n_{e}^{2}%
}{m^{3}}\right) \beta \exp -\left( \frac{\pi n_{e}\beta }{2m}\right) \right]
\eqnum{3.19}
\end{equation}
\begin{equation}
\xi _{2}\simeq e\sqrt{\frac{n_{e}}{m}}\left[ 1-\left( \frac{\pi n_{e}}{m}%
\right) \beta \exp -\left( \frac{\pi n_{e}\beta }{2m}\right) \right] 
\eqnum{3.20}
\end{equation}
These two inverse length scales correspond to two short-range
electromagnetic modes of propagation. These results are consistent with
those obtained in ref. \cite{8} using a different approach. If the masses of
the two massive modes, obtained in ref. \cite{8} by using the
electromagnetic thermal Green's function for static and slowly varying
configurations, are evaluated in the range of parameters considered above,
it can be shown that they reduce to eqs. (319), (3.20).

The solutions for $B$, $a_{0}$ and $A_{0}$, can be obtained using eqs.
(3.13), (3.14), (3.17) and the definition of $E$ in terms of $A_{0,}$

\begin{equation}
B\left( x\right) =\gamma _{1}\left( C_{2}e^{x\xi _{1}}-C_{1}e^{-x\xi
_{1}}\right) +\gamma _{2}\left( C_{4}e^{x\xi _{2}}-C_{3}e^{-x\xi
_{2}}\right) +C_{5}  \eqnum{3.21}
\end{equation}

\begin{equation}
a_{0}\left( x\right) =\chi \gamma _{1}\left( C_{2}e^{x\xi
_{1}}-C_{1}e^{-x\xi _{1}}\right) +\chi \gamma _{2}\left( C_{4}e^{x\xi
_{2}}-C_{3}e^{-x\xi _{2}}\right) +C_{6}  \eqnum{3.22}
\end{equation}

\begin{equation}
A_{0}\left( x\right) =\frac{1}{\xi _{1}}\left( C_{1}e^{-x\xi
_{1}}-C_{2}e^{x\xi _{1}}\right) +\frac{1}{\xi _{2}}\left( C_{3}e^{-x\xi
_{2}}-C_{4}e^{x\xi _{2}}\right) +C_{7}  \eqnum{3.23}
\end{equation}
In the above formulas we introduced the notation $\gamma _{1}=\left( \xi
_{1}^{2}\kappa +\eta \right) /\xi _{1}$, $\gamma _{2}=\left( \xi
_{2}^{2}\kappa +\eta \right) /\xi _{2}$.

In obtaining eq. (3.16) we have taken the derivative of eq. (3.12).
Therefore, the solution of eq. (3.16) belongs to a wider class than the one
corresponding to eqs. (3.12)-(3.14). To exclude redundant solutions we must
require that they satisfy eq. (3.12) as a supplementary condition. In this
way the number of independent unknown coefficients is reduced to six, which
is the number corresponding to the original system (3.12)-(3.14). The extra
unknown coefficient is eliminated substituting the solutions (3.17), (3.21),
(3.22) and (3.23) into eq. (3.12) to obtain the relation

\begin{equation}
e{\it \Pi }_{{\it 1}}C_{5}=-{\it \Pi }_{{\it 0}}\left( C_{6}+eC_{7}\right) 
\eqnum{3.24}
\end{equation}

Eq. (3.24) has an important meaning, it establishes a connection between the
coefficients of the long-range modes of the zero components of the gauge
potentials $(C_{6}+eC_{7})$ and the coefficient of the long-range mode of
the magnetic field $C_{5}$. Note that if the induced Chern-Simons
coefficient ${\it \Pi }_{{\it 1}}$, or the Debye screening coefficient ${\it %
\Pi }_{{\it 0}}$ were zero, there would be no link between $C_{5}$ and $%
(C_{6}+eC_{7})$. This relation between the long-range modes of $B$, $A_{0}$
and $a_{0}$ can be interpreted as a sort of Aharonov-Bohm effect, which
occurs in this system at finite temperature. At $T=0$, we have ${\it \Pi }_{%
{\it 0}}=0$, and the effect disappears.

Up to this point no boundary has been taken into account. Therefore, it is
easy to understand that the magnetic long-range mode associated with the
coefficient $C_{5}$, must be identified with the one found in ref. \cite{8}
for the infinite bulk using a different approach. However, as it is shown
below, when a constant and uniform magnetic field is perpendicularly applied
at the boundaries of a two-dimensional sample, this mode cannot propagate
(i.e. $C_{5}\equiv 0$) within the sample. This result is crucial for the
existence of Meissner effect in this system.

In order to determine the unknown coefficients we need to use the boundary
conditions. Henceforth we consider that the anyon fluid is confined to the
strip $-\infty <y<\infty $ with boundaries at $x=-L$ and $x=L$. The external
magnetic field will be applied from the vacuum at both boundaries ($-\infty
<x\leq -L$, $\;L\leq x<\infty $).

The boundary conditions for the magnetic field are $B\left( x=-L\right)
=B\left( x=L\right) =\overline{B}$ ($\overline{B}$ constant). Because no
external electric field is applied, the boundary conditions for this field
are, $E\left( x=-L\right) =E\left( x=L\right) =0$. Using them and assuming $%
L\gg \lambda _{1}$, $\lambda _{2}$ ($\lambda _{1}=1/\xi _{1}$, $\lambda
_{2}=1/\xi _{2}$), we find the following relations that give $C_{1,2,3,4}$
in terms of $C_{5}$,

\begin{equation}
C_{1}=Ce^{-L\xi _{1}},\quad C_{2}=-C_{1},\quad C_{3}=-Ce^{-L\xi _{2}},\quad
C_{4}=-C_{3},\quad C=\frac{C_{5}-\overline{B}}{\gamma _{1}-\gamma _{2}} 
\eqnum{3.25}
\end{equation}

\section{Stability Condition for the Infinite-Strip Sample}

After using the boundary conditions, we can see from (3.25) that they were
not sufficient to find the coefficient $C_{5}$. In order to totally
determine the system magnetic response we have to use another physical
condition from where $C_{5}$ can be found. Since, obviously, any meaningful
solution have to be stable, the natural additional condition to be
considered is the stability equation derived from the system free energy.
With this goal in mind we start from the free energy of the infinite-strip
sample

\[
{\cal F}=\frac{1}{2}\int\limits_{-L^{\prime }}^{L^{\prime
}}dy\int\limits_{-L}^{L}dx\left\{ \left( E^{2}+B^{2}\right) +\frac{N}{\pi }%
a_{0}b-{\it \Pi }_{{\it 0}}\left( eA_{0}+a_{0}\right) ^{2}\right. 
\]

\begin{equation}
\left. -{\it \Pi }_{{\it 0}}\,^{\prime }\left( eE+{\cal E}\right) ^{2}-2{\it %
\Pi }_{{\it 1}}\left( eA_{0}+a_{0}\right) \left( eB+b\right) +{\it \Pi }_{\,%
{\it 2}}\left( eB+b\right) ^{2}\right\}  \eqnum{4.1}
\end{equation}
where $L$ and $L^{\prime }$ determine the two sample's lengths.

Using the field solutions (3.17), (3.21)-(3.23) with coefficients (3.25), it
is found that the leading contribution to the free-energy density ${\it f}=%
\frac{{\cal F}}{{\cal A}}$ ,\ (${\cal A}=4LL^{\prime }$ being the sample
area) in the infinite-strip limit $(L\gg \lambda _{1}$, $\lambda _{2}$, $%
L^{\prime }\rightarrow \infty )$ is given by

\begin{equation}
f=C_{5}^{2}-{\it \Pi }_{{\it 0}}\left( C_{6}+eC_{7}\right) ^{2}+e^{2}{\it %
\Pi }_{\,{\it 2}}C_{5}^{2}-2e{\it \Pi }_{{\it 1}}\left( C_{6}+eC_{7}\right)
C_{5}  \eqnum{4.2}
\end{equation}

Taking into account the constraint equation (3.24), the free-energy density
(4.2) can be written as a quadratic function in $C_{5}$. Then, the value of $%
C_{5}$ is found, by minimizing the corresponding free-energy density

\begin{equation}
\frac{\delta {\it f}}{\delta C_{5}}=\left[ {\it \Pi }_{{\it 0}}+e^{2}{\it %
\Pi }_{{\it 1}}^{\,}\,^{2}+e^{2}{\it \Pi }_{{\it 0}}{\it \Pi }_{\,{\it 2}%
}\right] \frac{C_{5}}{{\it \Pi }_{{\it 0}}}=0,  \eqnum{4.3}
\end{equation}
to be $C_{5}=0$.

This result implies that the long-range mode cannot propagate within the
infinite-strip when a uniform and constant magnetic field is perpendicularly
applied at the sample's boundaries.

We want to point out the following fact. The same property of the finite
temperature polarization operator component $\Pi _{00}$ that is producing
the appearance of a long-range mode in the infinite bulk, is also
responsible, when it is combined with the boundary conditions, for the
non-propagation of this mode in the bounded sample. It is known that the
nonvanishing of ${\it \Pi }_{{\it 0}}$ at $T\neq 0$ (or equivalently, the
presence of a pole $\sim 1/k^{2}$ in $\Pi _{00}$ at $T\neq 0$) guarantees
the existence of a long-range mode in the infinite bulk \cite{8}. On the
other hand, however, once ${\it \Pi }_{{\it 0}}$ is different from zero, we
can use the constraint (3.24) to eliminate $C_{6}+eC_{7}$ in favor of $C_{5%
\text{ }}$ in the free-energy density of the infinite strip. Then, as we
have just proved, the only stable solution of this boundary-value problem,
which is in agreement with the boundary conditions, is $C_{5}=0$.
Consequently, no long-range mode propagates in the bounded sample.

In the zero temperature limit $\left( \beta \rightarrow \infty \right) $,
because ${\it \Pi }_{{\it 0}}=0$, it is straightforwardly obtained from
(3.24) that $C_{5}=0$ and no long-range mode propagates.

At $T\neq 0$, taking into account that $C_{5}=0$ along with eq. (3.25) in
the magnetic field solution (3.21), we can write the magnetic field
penetration as

\begin{equation}
B\left( x\right) =\overline{B}_{1}\left( T\right) \left( e^{-(x+L)\xi
_{1}}+e^{\left( x-L\right) \xi _{1}}\right) +\overline{B}_{2}\left( T\right)
\left( e^{-(x+L)\xi _{2}}+e^{\left( x-L\right) \xi _{2}}\right)  \eqnum{4.4}
\end{equation}
where,

\begin{equation}
\overline{B}_{1}\left( T\right) =\frac{\gamma _{1}}{\gamma _{1}-\gamma _{2}}%
\overline{B},\text{ \qquad \quad }\overline{B}_{2}\left( T\right) =\frac{%
\gamma _{2}}{\gamma _{2}-\gamma _{1}}\overline{B}  \eqnum{4.5}
\end{equation}

For densities $n_{e}\ll m^{2}$, the coefficients $\overline{B}_{1}$and $%
\overline{B}_{2}$ can be expressed, in the low temperature approximation $%
\left( T\ll \omega _{c}\right) $, as

\begin{equation}
B_{1}\left( T\right) \simeq -\frac{\left( \pi n_{e}\right) ^{3/2}}{m^{2}}%
\left[ 1/m+\frac{5}{2}\beta \exp -\left( \frac{\pi n_{e}\beta }{2m}\right)
\right] \overline{B},\qquad  \eqnum{4.6}
\end{equation}
\begin{equation}
B_{2}\left( T\right) \simeq \left[ 1+\frac{5\pi n_{e}}{2m^{2}}\sqrt{\pi n_{e}%
}\beta \exp -\left( \frac{\pi n_{e}\beta }{2m}\right) \right] \overline{B} 
\eqnum{4.7}
\end{equation}

Hence, in the infinite-strip sample the applied magnetic field is totally
screened within the anyon fluid on two different scales, $\lambda _{1}=1/\xi
_{1}$ and $\lambda _{2}=1/\xi _{2}$. At $T=200K$, for the density value
considered above, the penetration lengths are given by $\lambda _{1}\simeq
0.6\times 10^{-8}cm$ and $\lambda _{2}\simeq 0.1\times 10^{-4}cm$ .
Moreover, taking into account that $\xi _{1}$ increases with the temperature
while $\xi _{2}$ decreases (see eqs. (3.19)-(3.20)), and that $B_{1}\left(
T\right) <0$ while $B_{2}\left( T\right) >0$, it can be shown that the
effective penetration length $\overline{\lambda }$ (defined as the distance $%
x$ where the magnetic field falls down to a value $B\left( \overline{\lambda 
}\right) /\overline{B}=e^{-1}$) increases with the temperature as

\begin{equation}
\overline{\lambda }\simeq \overline{\lambda }_{0}\left( 1+\overline{\kappa }%
\beta \exp -\frac{1}{2}\overline{\kappa }\beta \right)   \eqnum{4.8}
\end{equation}
where $\overline{\lambda }_{0}=\sqrt{m/n_{e}e^{2}}$ and $\overline{\kappa }%
=\pi n_{e}/m$. At $T=200K$ the effective penetration length is $\overline{%
\lambda }\sim 10^{-5}cm$.

It is timely to note that the presence of explicit (proportional to $N$) and
induced (proportional to ${\it \Pi }_{{\it 1}}$) Chern-Simons terms in the
anyon effective action (3.1) is crucial to obtain the Meissner solution
(4.4). If the Chern-Simons interaction is disconnected ($N\rightarrow \infty 
$ and ${\it \Pi }_{{\it 1}}=0$), then $a=0,$ $d=1+e^{2}{\it \Pi }_{{\it 0}%
}{}^{\prime }\neq 0$ and $c=e^{2}{\it \Pi }_{{\it 0}}\,\neq 0$ in eq.
(3.16). In that case the solution of the field equations within the sample
are $E=0$, $B=\overline{B}$. That is, we regain the QED in (2+1)-dimensions,
which does not exhibit any superconducting behavior.

\section{High Temperature Non-Superconducting Phase}

We have just found that the charged anyon fluid confined to an infinite
strip exhibits Meissner effect at temperatures lower than the energy gap $%
\omega _{c}$. It is natural to expect that at temperatures larger than the
energy gap this superconducting behavior should not exist. At those
temperatures the electron thermal fluctuations should make available the
free states existing beyond the energy gap. As a consequence, the charged
anyon fluid should not be a perfect conductor at $T\gg \omega _{c}$. A
signal of such a transition can be found studying the magnetic response of
the system at those temperatures.

As can be seen from the magnetic field solution (4.4), the real character of
the inverse length scales (3.18) is crucial for the realization of the
Meissner effect. At temperatures much lower than the energy gap this is
indeed the case, as can be seen from eqs. (3.19) and (3.20).

In the high temperature $\left( T\gg \omega _{c}\right) $ region the
polarization operator coefficients are given by eq. (3.9). Using this
approximation together with the assumption $n_{e}\ll m^{2}$, we can
calculate the coefficients $a$, $c$ and $d$ that define the behavior of the
inverse length scales,

\begin{equation}
a\simeq \pi ^{2}{\it \Pi }_{{\it 0}}{}^{\prime }{\it \Pi }_{\,{\it 2}} 
\eqnum{5.1}
\end{equation}

\begin{equation}
c\simeq e^{2}{\it \Pi }_{{\it 0}}{}  \eqnum{5.2}
\end{equation}

\begin{equation}
d\simeq -1  \eqnum{5.3}
\end{equation}

Substituting with (5.1)-(5-3) in eq. (3.18) we obtain that the inverse
length scales in the high-temperature limit are given by

\begin{equation}
\xi _{1}\simeq e\sqrt{m/2\pi }\left( \tanh \frac{\beta \mu }{2}+1\right) ^{%
\frac{1}{2}}  \eqnum{5.4}
\end{equation}

\begin{equation}
\xi _{2}\simeq i\left[ 24\sqrt{\frac{2m}{\beta }}\cosh \frac{\beta \mu }{2}%
\left( \tanh \frac{\beta \mu }{2}+1\right) ^{-\frac{1}{2}}\right]  
\eqnum{5.5}
\end{equation}

The fact that $\xi _{2}$ becomes imaginary at temperatures larger than the
energy gap, $\omega _{c}$, implies that the term $\gamma _{2}\left(
C_{4}e^{x\xi _{2}}-C_{3}e^{-x\xi _{2}}\right) $ in the magnetic field
solution (3.21) ceases to have a damping behavior, giving rise to a periodic
inhomogeneous penetration. Therefore, the fluid does not exhibit a Meissner
effect at those temperatures since the magnetic field will not be totally
screened. This corroborate our initial hypothesis that at $T\gg \omega _{c}$
the anyon fluid is in a new phase in which the magnetic field can penetrate
the sample.

We expect that a critical temperature of the order of the energy gap ($T\sim
\omega _{c}$) separates the superconducting phase $\left( T\ll \omega
_{c}\right) $ from the non-superconducting one $\left( T\gg \omega
_{c}\right) $. Nevertheless, the temperature approximations (3.8) and (3.9)
are not suitable to perform the calculation needed to find the phase
transition temperature. The field solutions in this new non-superconducting
phase is currently under investigation. The results will be published
elsewhere.

\section{Concluding Remarks}

In this paper we have investigated the magnetic response at finite
temperature of a charged anyon fluid confined to an infinite strip. The
charged anyon fluid was modeled by a (2+1)-dimensional MCS theory in a
many-particle ($\mu \neq 0$, $\overline{b}\neq 0$) ground state. The
particle energy spectrum of the theory exhibits a band structure given by
different Landau levels separated by an energy gap $\omega _{c}$, which is
proportional to the background Chern-Simons magnetic field $\overline{b}$.
We found that the energy gap $\omega _{c}$ defines a scale that separates
two phases: a superconducting phase at $T\ll \omega _{c}$, and a
non-superconducting one at $T\gg \omega _{c}$.

The total magnetic screening in the superconducting phase is characterized
by two penetration lengths corresponding to two short-range eigenmodes of
propagation of the electromagnetic field within the anyon fluid. The
existence of a Meissner effect at finite temperature is the consequence of
the fact that a third electromagnetic mode, of a long-range nature, which is
present at finite temperature in the infinite bulk \cite{8}, does not
propagate within the infinite strip when a uniform and constant magnetic
field is applied at the boundaries. This is a significant property since the
samples used to test the Meissner effect in high-$T_{c}$ superconductors are
bounded.

It is noteworthy that the existence at finite temperature of a Debye
screening (${\it \Pi }_{{\it 0}}\,\neq 0$) gives rise to a sort of
Aharonov-Bohm effect in this system with Chern-Simons interaction ($N$
finite, ${\it \Pi }_{{\it 1}}\neq 0$). When ${\it \Pi }_{{\it 0}}\,\neq 0$,
the field combination $a_{0}+eA_{0}$ becomes physical because it enters in
the field equations in the same foot as the electric and magnetic fields
(see eq. (3.12)). A direct consequence of this fact is that the coefficient $%
C_{5}$, associated to the long-range mode of the magnetic field, is linked
to the coefficients $C_{6}$ and $C_{7}$ of the zero components of the
potentials (see eq. (3.24)).

When $T=0$, since ${\it \Pi }_{{\it 0}}\,=0$ and ${\it \Pi }_{{\it 1}}\neq 0$%
, eq. (3.24) implies $C_{5}=0$. That is, at zero temperature the long-range
mode is absent. This is the well known Meissner effect of the anyon fluid at 
$T=0$. When $T\neq 0$, eq. (3.24) alone is not enough to determine the value
of $C_{5}$, since it is given in terms of $C_{6}$ and $C_{7}$ which are
unknown. However, when eq. (3.24) is taken together with the field
configurations that satisfy the boundary conditions for the infinite-strip
sample (eqs. (3.17), (3.21)-(3.23) and (3.25)), and with the sample
stability condition (4.3), we obtain that $C_{5}=0$. Thus, the combined
action of the boundary conditions and the Aharonov-Bohm effect expressed by
eq. (3.24) accounts for the total screening of the magnetic field in the
anyon fluid at finite temperature.

Finally, at temperatures large enough ($T\gg \omega _{c}$) to excite the
electrons beyond the energy gap, we found that the superconducting behavior
of the anyon fluid is lost. This result was achieved studying the nature of
the characteristic lengths (3.18) in this high temperature approximation. We
showed that in this temperature region the characteristic length $\xi _{2}$
becomes imaginary (eq. (5.5)), which means that a total damping solution for
the magnetic field does not exist any more, and hence the magnetic field
penetrates the sample.

\begin{quote}
Acknowledgments
\end{quote}

The authors are very grateful for stimulating discussions with Profs. G.
Baskaran, A. Cabo, E.S. Fradkin, Y. Hosotani and J. Strathdee. We would also
like to thank Prof. S. Randjbar-Daemi for kindly bringing the publication of
ref. \cite{b} to our attention. Finally, it is a pleasure for us to thank
Prof. Yu Lu for his hospitality during our stay at the ICTP, where part of
this work was done. This research has been supported in part by the National
Science Foundation under Grant No. PHY-9414509.

\end{document}